\documentclass[journal,12pt,onecolumn]{IEEEtran}
\renewcommand{\baselinestretch}{1.7}
\usepackage{cite}

\pdfoutput=1 

\usepackage[pdftex]{graphicx}

\usepackage{amsmath}

\usepackage{algorithmic}

\usepackage{array}

\ifCLASSOPTIONcompsoc
 \usepackage[caption=false,font=normalsize,labelfont=sf,textfont=sf]{subfig}
\else
 \usepackage[caption=false,font=footnotesize]{subfig}
 \fi
\usepackage{amssymb}  
\DeclareMathAlphabet\mathbfcal{OMS}{cmsy}{b}{n}
\usepackage{color}
\usepackage{epsfig}

\usepackage{amssymb}
\begin{document}
\title{Reconfigurable Antennas in mmWave MIMO Systems}

\author{Mojtaba~Ahmadi~Almasi,~\IEEEmembership{Student~Member,~IEEE,}
        Hani~Mehrpouyan,~\IEEEmembership{Member,~IEEE,} 
        Vida~Vakilian,~\IEEEmembership{Member,~IEEE,} 
        Nader~Behdad,~\IEEEmembership{Fellow,~IEEE,}
        and~Hamid~Jafarkhani,~\IEEEmembership{Fellow,~IEEE}

\thanks{M. A. Almasi and H. Mehrpouyan are with the Department of Electrical and Computer Engineering, Boise State University, Boise, ID 83725 USA (e-mails: mojtabaahmadialm@boisestate.edu, hani.mehr@ieee.org). V. Vakilian is with the Department of Electrical and Computer Engineering and Computer Science, California State University, Bakersfield, CA 93311 USA (e-mail: vvakilian@csub.edu). N. Behdad is with the Department of Electrical and Computer Engineering, University of Wisconsin$-$Madison, Madison, WI 30332 USA (e-mail: behdad@wisc.edu). H. Jafarkhani is with the Center for Pervasive Communications and Computing, University of California, Irvine, CA 92697 USA (e-mail: hamidj@uci.edu).

This project is supported in part by the NSF ERAS grant award numbers 1642865, 1642536, 1642601, 1642567.}
}
\markboth{}%
{Shell \MakeLowercase{\textit{et al.}}: Bare Demo of IEEEtran.cls for IEEE Journals}
\maketitle

\section*{Abstract}
The key obstacle to achieving the full potential of the millimeter wave (mmWave) band has been the poor propagation characteristics of wireless signals in this band. One approach to overcome this issue is to use antennas that can support higher gains while providing beam adaptability and diversity, i.e., reconfigurable antennas. In this article, we present a new architecture for mmWave multiple-input multiple-output (MIMO) communications that uses a new class of reconfigurable antennas. More specifically, the proposed lens-based antennas can support multiple radiation patterns while using a single radio frequency chain. Moreover, by using a beam selection network, each antenna beam can be steered in the desired direction. Further, using the proposed reconfigurable antenna in a MIMO architecture, we propose a new signal processing algorithm that uses the additional degrees of freedom provided by the antennas to overcome propagation issues at mmWave frequencies. Our simulation results show that the proposed reconfigurable antenna MIMO architecture significantly enhances the performance of mmWave communication systems.

\section{Introduction}
Wireless systems are increasingly supporting larger and more diverse applications from sensor networks for environmental monitoring, to smart grid electrical infrastructures, to advances in medicine and transportation. To meet this demand, cellular providers need to have access to more bandwidth, which is their primary capital expenditure. They could reduce such costs$-$and introduce potentially far reaching improvements to cellular access, affordability, and coverage$-$by making better use of available spectrum in the $30$$-$$300$ GHz millimeter-wave (mmWave) band\textcolor{black}{\cite{r1}}. However, propagation and hardware challenges, such as large path loss \textcolor{black}{\cite{r2}}, severe shadowing \textcolor{black}{\cite{r2}}, amplifier limitations \textcolor{black}{\cite{r2}}, phase noise \textcolor{black}{\cite{r4}}, and large power consumption by high speed digital signal processing units (digital-to-analog and analog-to-digital converters (DAC/ADC)) \textcolor{black}{\cite{r2}}, have prevented this. 

To address the propagation issues outlined above, several architectures have been proposed for mmWave communication systems, which are different from that of systems operating in the sub-$6$ GHz frequencies~\cite{r13}. In one approach researchers have proposed the use of massive antenna arrays at the transmitter and receiver sides to mitigate large path loss at mmWave frequencies. The short wavelengths at mmWave frequencies allows for the deployment of such large antenna arrays. However, the deployment of these massive multi-input multi-output (MIMO) systems, brings about two main implementation challenges: (1) Equipping each antenna with a radio frequency (RF) chain at mmWave frequencies is costly, extremely complex, and space limiting. (2) Such an architecture will result in large power consumption due to the massive number of RF chains (consisting of power amplifiers, DAC/ADCs etc.) that need to be used\footnote{The larger bandwidth used at mmWave frequencies requires the use of high frequency ADC/DACs, which are very power hungry.}. It is important to note that aside from the above implementation issues, massive MIMO systems further suffer from shadowing and channel sparsity, which degrade their performance at mmWave frequencies. To address these challenges, here, we proposed the use of reconfigurable antennas. 

Most communication systems to date use traditional antennas with static radiation patterns. These antennas cannot modify their radiation patterns and frequency of operation, which can be a severe limitation for applications in future wireless networks that are expected to operate both in the sub $6$ GHz and also at mmWave frequencies. This is needed since mmWave communication systems suffer from significant propagation issues, e.g., large path loss and shadowing, that require more directional and adaptive radiation patterns. Hence, researchers have proposed a class of antennas, denoted by, reconfigurable antennas, which can dynamically change their radiation characteristics. Our current reconfigurable antenna design and prior work in the field have shown that reconfigurable antennas can provide radiation pattern diversity, which we show can be used to overcome propagation issues at mmWave frequencies.

 
To reliably meet the needs of a massive user base and potentially improve access for the underserved to wireless-enabled goods and services, this article aims to develop a communication theory, and antenna design framework for reconfigurable antenna systems to overcome these challenges. Instead of using a large array of static antennas, our proposed approach uses a small number of newly proposed reconfigurable antennas. These antennas can support simultaneous transmission of multiple radiation patterns with one RF chain. These additional radiation patterns provide a communication system with more degrees of freedom. Hence, aside from being more cost effective, the proposed methodology allows for more effective beamforming and beamsteering algorithms that can better overcome propagation issues in the mmWave band. Further, the large gain provided by the proposed reconfigurable antenna design reduces the mmWave power amplifier design constraints. Using the proposed antenna in a MIMO architecture, we also propose a new signal processing algorithm that uses the additional degrees of freedom provided by the antennas to overcome propagation issues at mmWave frequencies. Our simulation results show that the proposed reconfigurable antenna MIMO architecture significantly enhances the performance of MIMO mmWave communication systems.

\section{Motivation} 
Here, we outline the potential value of using a small number of reconfigurable antennas with various known radiation states to overcome the propagation
obstacles at mmWave frequencies. 

\begin{figure}
\includegraphics[scale = 0.9]{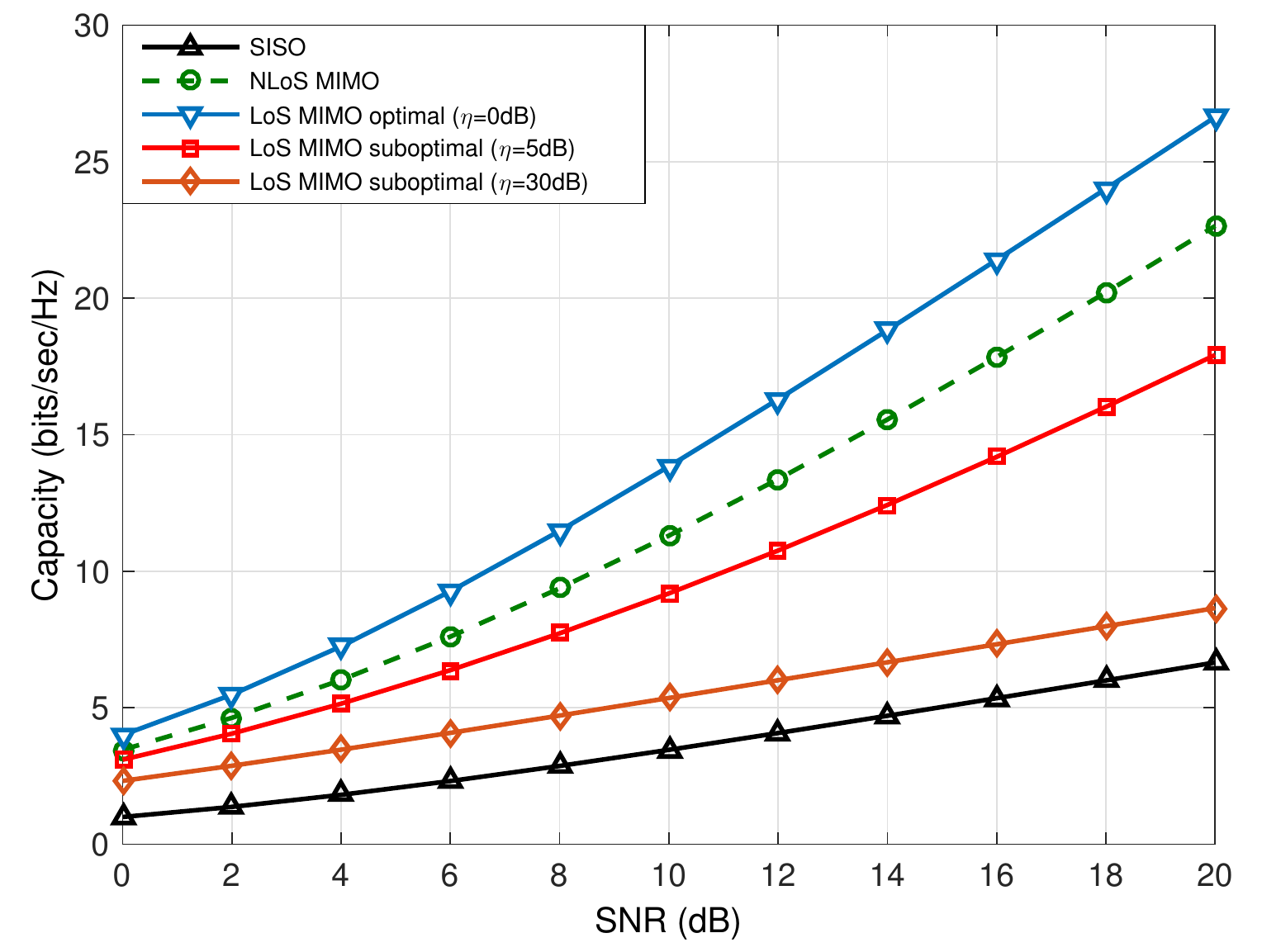}
\centering
\caption{Capacity of a $4 \times 4$ MIMO system vs. the deviation with respect to optimal transceiver and antenna spacing measured by $\eta = \frac{\text{Antenna}\text{ }\text{Seperation}\text{ }\text{Product}_{\text{opt}}}{\text{Antenna}\text{ }\text{Seperation}\text{ }\text{Product}}$}
 \label{fig1}
\end{figure}

MmWave communications mainly takes place through line-of-sight (LoS) links to overcome the large signal attenuation in this band~\cite{r2 , r4} and to take advantage of the greater antenna directivity at these higher frequencies. Researchers have also demonstrated that MIMO systems combined with beamforming approaches are likely to further circumvent the path loss issues at mmWave frequencies \textcolor{black}{\cite{r6}}. However, even with the application of the above solutions, the resulting LoS MIMO channels are known to be low-rank and/or sparse. This is because LoS MIMO systems can only achieve full-rank channels when the transmitter and receiver distance and the antenna spacings are set to specific optimal values~\cite{r8}. As shown in Fig. \ref{fig1}, deviations from this optimal spacing (denoted by $\eta$) can result in significant performance degradation. 

One frequently explored solution is to use a large array of antennas in the mmWave band to obtain gains similar to those of small none line-of-sight (NLoS) MIMO systems that take advantage of rich scattering. However, this approach can add complexity, costs, and potentially large loss. As a more applicable and fundamentally different approach, we propose to use reconfigurable antennas and their various predefined states to modify the radiation pattern of each element of the MIMO array and to eliminate the dependency on the optimal antenna and transceiver spacing. Note that the relationship between each antenna state and the resulting radiation pattern is known and determined at the design stage~\cite{r5}. Moreover, the various known states of reconfigurable antennas allow for new beamforming and beamsteering approaches for MIMO systems, which design engineers can in turn use to more effectively tackle significant shadowing, path loss, and channel sparsity at the mmWave band. 

Recall that the main characteristics of mmWave communications are short wavelength, large bandwidth, and high attenuation. These characteristics lead to a sparse-scattering environment, where the majority of the directions of arrivals are below the noise floor. This results in potentially sparse and/or ill-conditioned mmWave MIMO channels.

Channel sparsity is known to limit the performance of LoS mmWave MIMO systems (see Fig. \ref{fig1}). Sayed et al. have perhaps completed the most consequential work to date attempting to address this issue~\cite{r15}. This work shows that it is feasible to circumvent channel sparsity in MIMO systems by varying the spacing amongst the antennas in an array, resulting in significant performance gains. Nevertheless, this work only focuses on modifying the antenna spacings in an array, which results in limited degrees of freedom. In contrast, by using the proposed microelectromechanical systems (MEMs), PIN diodes, and passive/active elements, each reconfigurable antenna has a larger number of known states that when combined in a MIMO configuration can result in a significantly larger degrees of freedom. In fact, aside from our work in (U.S. Patent 15,284,123), there are very few results to date that utilize the degrees of freedom provided by reconfigurable antennas to address channel sparsity in MIMO mmWave systems.

\begin{figure}
\includegraphics[scale = 0.6]{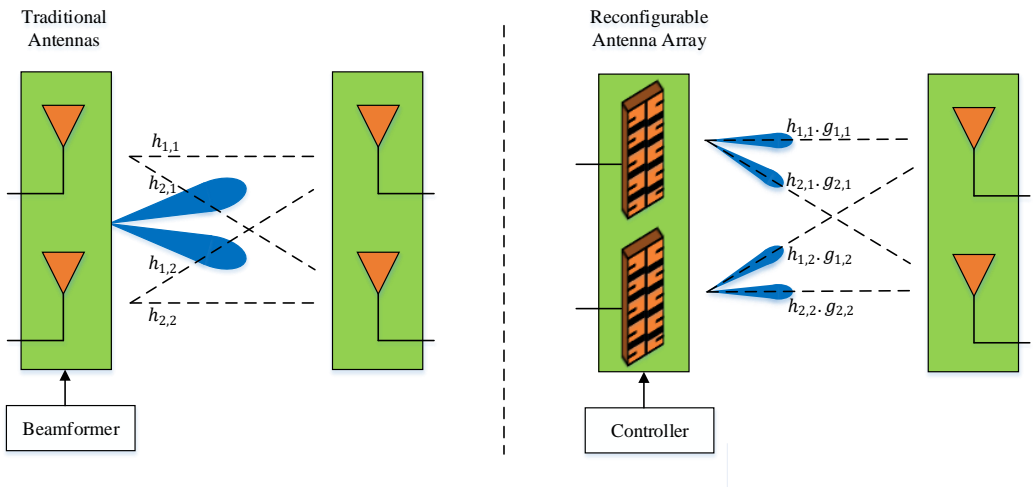}
\centering
\caption{A $2 \times 2$ MIMO system equipped with reconfigurable antennas at the transmitter side can form four main radiation lobes, while one equipped with conventional antennas can at most form two main radiation lobes via beamforming. $h_{i,j}$ for $i,j = 1$ and $2$ represents the channel coefficient between the $i$th receive antenna and $j$th transmit antenna. Also, $g_{i,j}$ for $i,j = 1$ and $2$ denotes the reconfigurable antenna state of the $j$th radiation lobes of the $i$th reconfigurable antenna.}
 \label{fig2}
\end{figure}
Fig. \ref{fig2} demonstrates the proposed concept and compares it with beamforming for traditional antennas. As shown in Fig. \ref{fig2}, a 2 $\times$ 2 MIMO system equipped with reconfigurable antennas can support more radiation patterns compared to that of a system that is equipped with static antennas. Further, our preliminary results, see (U.S. Patent 15,284,123) and \cite{r3}, show that we can design reconfigurable antennas that not only support multiple radiation patterns but each radiation pattern from an antenna can be \textit{independently} steered. However, to date, there is a lack of a signal processing and an antenna theory framework for using of these antennas at mmWave frequencies. In this article, we intend to address this shortcoming by proposing a new reconfigurable antenna design and a new signal processing algorithm that uses the characteristics of these antennas in mmWave MIMO systems.  

\section{Proposed Reconfigurable Antenna Design for Millimeter Wave Communications}
Mitigation of sparse and low-rank mmWave channels requires the development of reconfigurable antennas that can support the simultaneous transmission of multiple radiation patterns. Therefore, in this section we propose a reconfigurable antenna design with these specific characteristics.  

\begin{figure}
\includegraphics[scale = 1]{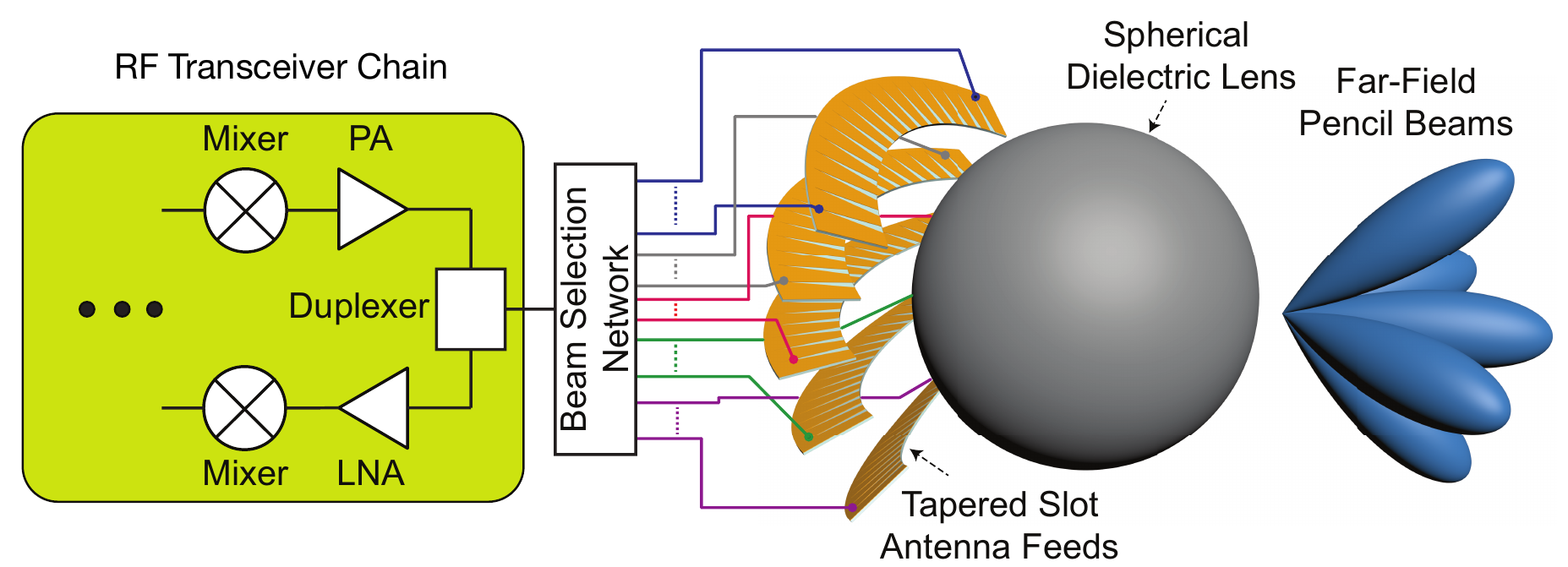}
\centering
\caption{Topology of the proposed multi-beam, reconfigurable antenna. The antenna is composed of a spherical lens fed with a number of tapered slot antenna feeds. Each feed generates a beam in a given direction in the far field.}
 \label{fig3}
\end{figure}
To accomplish this, we have used a lens-based multi-beam antenna similar to the one shown in Fig.~\ref{fig3}. This structure is composed of a spherical dielectric lens fed with multiple tapered slot antennas (TSAs). The combination of each TSA and the spherical lens produces a highly directive radiation pattern in the far field. Each TSA feed generates a beam in a given direction in the far field. Therefore, this system is a multi-beam antenna capable of generating $N_f$ independent beams where $N_{f}$ is the number of TSA feeds. While this antenna uses many different feeds, only the feed antennas that generate the beams in the desired directions need to be excited. As a result, this antenna system only uses one transceiver chain to generate the desired number of radiated beams. This scenario is shown in Fig.~\ref{fig3} where the output of the transceiver chain is connected to a $``$Beam Selection Network$"$. This sub-system has $N_f$ outputs that are connected to the input ports of the TSA feeds. In the simplest scenario where only a single output beam is desired, the beam selector can be a simple single pole $N_f$ throw switch which connects the transceiver chain to the TSA feed antenna, and generates a beam in the desired direction. Alternatively, if the mmWave system design requires a radiation pattern with two (or $M > 2$) independently directed beams, we can design the beam selector using a single-pole-$N_f$-throw switch. This allows us to support a MIMO architecture with one RF chain at the transceiver.

\begin{figure}
\includegraphics[scale = 1.4]{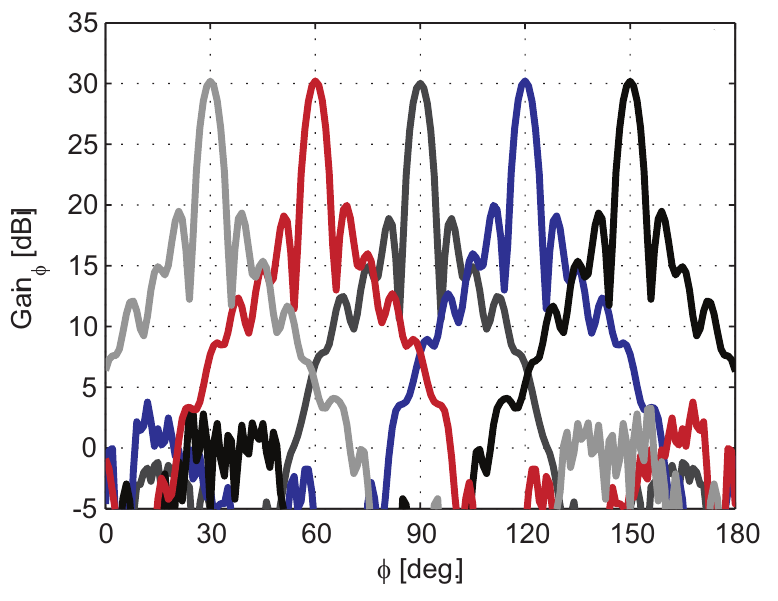}
\centering
\caption{Full-wave EM simulation results (from FEKO) of the radiation patterns of a lens antenna using five TSA feeds.}
 \label{fig4}
\end{figure}

\begin{figure}
\includegraphics[scale = 3]{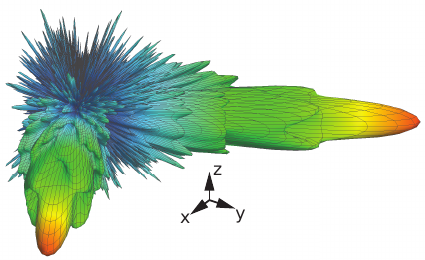}
\centering
\caption{Full-wave EM simulation results of a lens antenna with five TSA feeds when two of the feeds are simultaneously excited.}
 \label{fig5}
\end{figure}
Our new reconfigurable antenna design operates in the $50$-$70$ GHz frequency band and has five tapered slot antenna feeds. In this design, we used a simple Teflon sphere with a diameter of $65$ mm and a dielectric constant of $2.2$ as the lens and individual TSAs with dimensions of $25$ mm $\times$ $5$ mm as the feed. The lens and its feeds are optimized together, using full-wave EM simulations in FEKO, to maximize the far field gain of the antenna. Fig. \ref{fig4} shows the far-field radiation patterns that the feeds generate for this antenna. Observe that each feed generates a beam with a gain of $\sim$ $30$ dBi and side-lobe-levels below $12$ dB. Moreover, the design had enabled us to obtain an extremely wide scanning range spanning an angular range of $120$ degrees. This range can be easily extended to cover an angular range of $180$ degrees. Because of the rotational symmetry of the lens and its feeds, the radiation pattern of the antenna does not deteriorate with scanning angle as is common in planar antenna apertures. As described earlier, this antenna can generate a pattern with multiple independently controlled main lobes using a single RF transceiver chain with the aid of the $``$Beam Selector$"$ device shown in Fig.~\ref{fig3}. Fig.~\ref{fig5} shows the full-wave EM simulation results of such a scenario where the beam selector feeds the input signal to two of the five feed antennas in phase. Specifically, the beam selector is configured to generate two main lobes along the directions of $30$ and $120$ degrees in the azimuth plane. Both beams maintain their inherent gain of $30$ dBi as expected. The proposed reconfigurable antenna has the following unique attributes:
\begin{itemize}
\item \textbf{Independently steerable beams:} Using a single RF chain and an appropriate $``$Beam Selection$"$ network, a system using this reconfigurable antenna can form two or more independently steerable beams. This increases the degrees of freedom within the system and introduces additional diversity that can be used to overcome shadowing at mmWave frequencies via beam diversity. This approach also allows us to support an $N_f \times N_f$ diversity based MIMO setup by using a single reconfigurable antenna and by steering its radiation pattern in different directions instead of having multiple separate antennas and RF chains at both the transmitter and receiver. Although spatial multiplexing cannot be supported via one RF chain, bandwidth efficiency is not as important as link reliability at mmWave frequencies due to the availability of large unused bandwidth. In the next section, we will also show that $\mathbf{G}$, the reconfigurable antenna gain matrix, plays a crucial role in overcoming low-rank MIMO channels at mmWave frequencies. 

\item \textbf{Full hemispherical coverage:} Antenna beams can cover the entire hemispherical volume and provide 2$\pi$ steradian of coverage. Specifically, beam steering in both the azimuth and the elevation planes is possible using this approach. This allows for effective beam steering around objects that can severely block or shadow the signals aimed from the transmitter to the receiver at mmWave frequencies.
\end{itemize}
\section{Capacity of Reconfigurable MIMO Systems}
Here, we evaluate the capacity of the proposed architecture in two cases. The first case is a traditional $-$ sub-6 GHz $-$ MIMO system with \textit{M} and \textit{N} static omnidirectional antennas at the transmitter and  receiver sides, respectively. The second scenario is a mmWave MIMO system that is equipped with \textit{M} reconfigurable antennas at the transmitter and \textit{N} static omnidirectional antennas at the receiver.  Further, each transmit reconfigurable antenna radiates $N_f = N$ independent beams. 
Recall that in mmWave systems, the communication channel is mainly established through LoS links. As a result, here, we use the Rician fading channel model. A Rician channel is comprised of a LoS, $\mathbf{H}_{\text{LoS}}$, and a none LoS (NLOS), $\mathbf{H}_{\text{NLoS}},$ component, i.e., $\mathbf{H}=\sqrt{\frac{K}{K+1}}\mathbf{H}_{\text{LoS}}+\sqrt{\frac{1}{K+1}}\mathbf{H}_{\text{NLoS}}$ where $K$ repserents the Rician factor. The former corresponds to the deterministic component of the channel while the latter is used to present the scattering, which is random and is characterized by the Rayleigh fading channel model. Since in mobile scenarios optimal antenna and transceiver spacing for a LoS MIMO system cannot be achieved and as such, $\mathbf{H}_{\text{LoS}}$ will be a low rank matrix.  In such a practical case, by increasing the Rician factor, \textit{K}, the capacity of the MIMO channel decreases~\cite{r7}. However, by using reconfigurable antennas and the degrees of freedom they provide, we can design the antenna states in such a way that the reconfigured channel matrix $\mathbf{H}_{g}$ is always full rank irrespective of the antenna spacing. This matrix can be expressed as 
\begin{equation} \label{eq1}
\mathbf{H}_g = \mathbf{H}\circ \mathbf{G},
\end{equation}
where $\mathbf{H}$ of size $N\times M$ is the described Rician channel matrix and $\mathbf{G}$ of size $N\times M$ is the reconfigurable antenna state matrix. Also, $\circ$ stands for the Hadamard product which indicates there is a one-to-one mapping from an antenna state to the associated channel coefficient as shown in Fig.~\ref{fig2}.  Thus, the entries of the reconfigured channel matrix are denoted by $h_{g,i,j} = h_{i,j}g_{i,j}$ for $i = 1,~2,~\dots,~N$ and $j = 1,~2,~\dots,~M$; moreover, $h_{i,j}$ and $g_{i,j}$ are the entries of the channel matrix and the reconfigurable antenna state matrix, respectively. The reconfigured matrix $\mathbf{H}_g$ represents our new signal processing approach to overcome propagation issues at mmWave frequencies. 

Regarding the matrix $\mathbf{H}_g$, the capacity of a reconfigurable antenna MIMO system in bits/sec/Hz is defined as 
\begin{equation} \label{eq2}
C = \text{log}\Big(\text{det}\big(\mathbf{I}_N + \frac{\rho}{M}\mathbf{H}_g\mathbf{H}_g^H\big)\Big),
\end{equation}
where $\mathbf{I}_N$ is an $N\times N$ identity matrix. Also, $\rho$ is the transmitted SNR and the superscript $^H$ denotes the Hermitian transpose. The capacity expression for the traditional MIMO is obtained by replacing $\mathbf{H}_g$ with $\mathbf{H}$ in (\ref{eq2}) \cite{r16}. Using the proposed reconfigurable antennas, we numerically investigate the capacity of a $2\times 2$ mmWave MIMO system when the system is equipped with two antennas at the transmitter and receiver. In this setup power is equally allocated at the transmitter side. Also, we design the reconfigurable antennas state matrix, $\mathbf{G}$, as 
\begin{equation} \label{eq3}
\mathbf{G} = 
\begin{pmatrix}
1 & 1\\
1 & e^{j\pi}
\end{pmatrix},
\end{equation} 
where $j = \sqrt{-1}$. The above equation completes our signal processing approach to tackle the rank deficient issue at LoS channels. While the $2\times 2$ MIMO channel is low rank, the entry $g_{2,2} = e^{j\pi}$ of $\mathbf{G}$ enhances the value of $h_{g,2,2}$ in (\ref{eq1}) in a way to ensure that the matrix $\mathbf{H}_g$ becomes full rank and the expression in (\ref{eq2}) achieves higher capacity. Note that the other channel coefficients remain unchanged. Also, when the channel $\mathbf{H}$ is intrinsically full rank, the matrix $\mathbf{G}$ does not effect the rank. Therefore, utilizing the reconfigurable antenna and an appropriate signal processing approach can enhance the performance of the capacity.    
\begin{figure}
     \includegraphics[scale = 0.8]{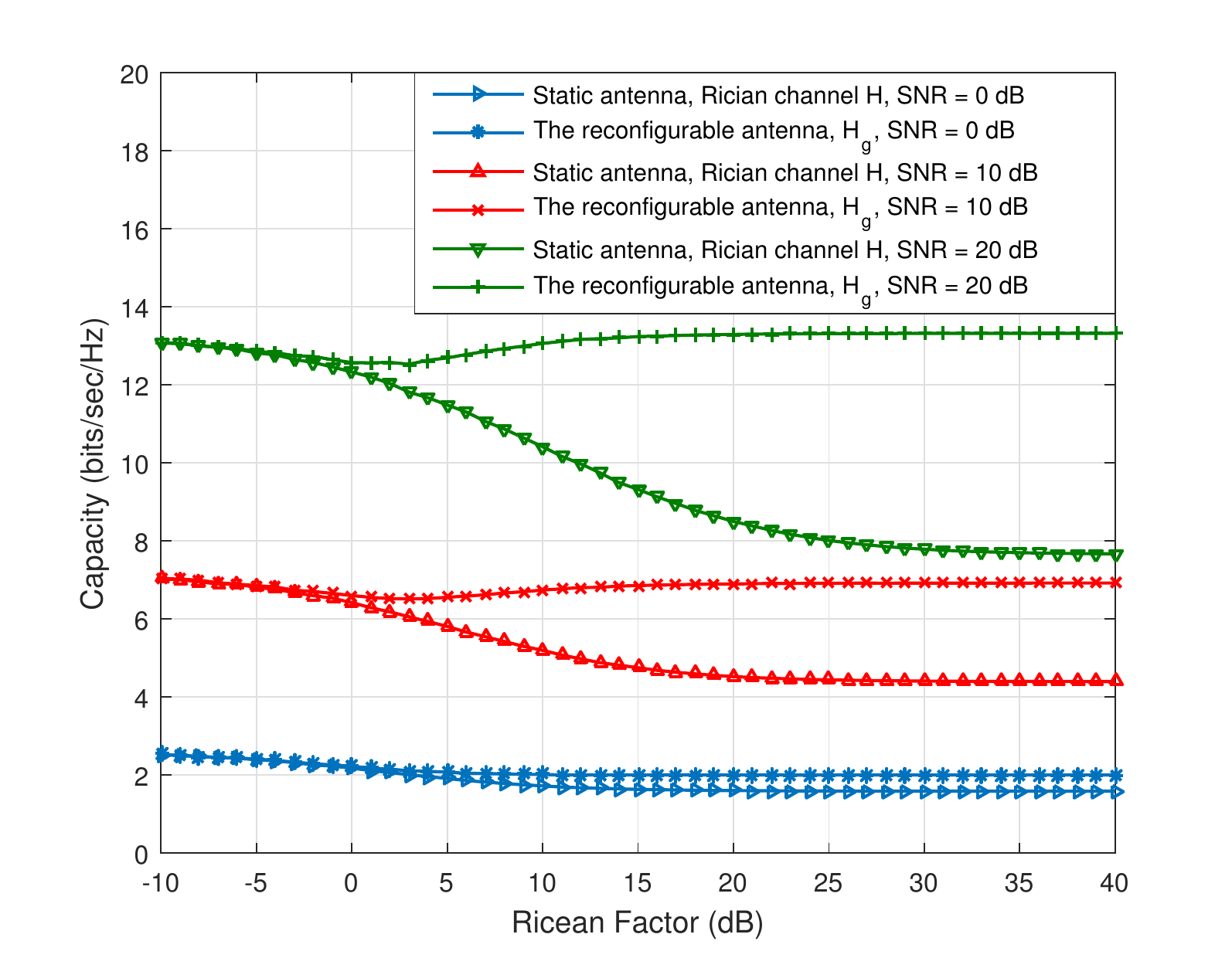}
     \centering
\caption{Capacity vs. the Rician factor of static and reconfigurable $2 \times 2$ MIMO systems. }
 \label{fig6}
\end{figure}


In Fig.~\ref{fig6}, as expected, the capacity of the traditional MIMO system decreases as the Rician factor increases. When channel exhibits a NLoS characteristic, i.e., low \textit{K} factor, there is enough scattering to ensure the MIMO channel is full rank, in turn ensuring a higher capacity. However, as the channel becomes more LoS and more closely exhibits the mmWave channel environment, the resulting scattering decreases and the MIMO channel becomes rank deficient, which in turn results in a lower capacity for traditional mmWave MIMO systems. Fig. \ref{fig6} also depicts the performance of a MIMO system that is equipped with reconfigurable antennas. By applying the degrees of freedom provided by the reconfigurable antennas, the proposed signal processing approach in (\ref{eq1}) and (\ref{eq3}) can ensure that the LoS component of the channel model, $\mathbf{H}_{\text{LoS}}$, is always full rank. This ensures that in mmWave settings where LoS links are predominant, i.e., a high Rician factor, the proposed system can deliver significantly better performance than traditional MIMO systems.

\section{Conclusion}
Combining the advantages of reconfigurable antennas with the potential of mmWave systems will result in significant capacity gains for future wireless systems. To this end, this article presents the potential of reconfigurable antennas in addressing the propagation issues at mmWave frequencies and outlines a series of open research problems for the application of reconfigurable antennas in mmWave systems. In the latter case, we also provide preliminary solutions for these open problems from both a signal processing and antenna theory perspective. From a signal processing perspective, we demonstrated that the additional antenna gain and degrees of freedom provided by the reconfigurable antennas can be used to overcome significant path loss and shadowing at mmWave frequencies. Moreover, by using the advantages of reconfigurable antennas, we can convert a low-rank mmWave LoS MIMO system channel into a full rank channel that can support multiple data streams, which in turn significantly enhances bandwidth efficiency and throughput gains. In addition, a new reconfigurable antenna design was proposed. The combination of tapered slot antennas and the spherical lens leads to designing an antennas with steerable and highly directive radiation patterns. Finally, we evaluated the effect of reconfigurable antennas with steerable radiation lobes on the performance of mmWave systems. By using this unique property of the proposed reconfigurable antennas, the rank of a LoS channel is improved leading to a capacity increase in mmWave systems. 


\bibliographystyle{IEEEtran}
\bibliography{IEEEabrv,references}
\end{document}